\documentclass[twoside,twocolumn,leqno]{article}
\usepackage{ltexpprt}

\usepackage{cite}
\usepackage{amsmath,amssymb,amsfonts}
\usepackage{algorithmic}
\usepackage{graphicx}
\usepackage{textcomp}
\usepackage{xcolor}

\usepackage{ulem}
\usepackage{etex}
\usepackage{subfigure}
\usepackage{graphicx}
\usepackage{amsmath}
\usepackage{bm}
\usepackage{url}
\usepackage{stmaryrd}
\usepackage{balance}
\usepackage{amssymb}
\usepackage{pifont}
\usepackage[usenames,dvipsnames]{pstricks}
\usepackage{epsfig}
\usepackage{epstopdf}
\usepackage{multirow}
\usepackage{booktabs}
\usepackage{pgffor}
\usepackage{tikz}
\usepackage{enumitem}
\usepackage{algorithm,algorithmic}
\usepackage{balance}
\usepackage{flushend}
\usepackage{acronym}
\graphicspath{{./figures/}} 
\usepackage{color}

\newcommand{\myred}[1]{\textcolor{red}{#1}}

\newcommand{\ours}[0]{$\text{Deep Program Reidentification}$}
\newcommand{\ourss}[0]{$\textbf{DeepHGNN}$}
\newcommand{\nop}[1]{}
\newcommand{\problem}{\textit{program reidentification}}

\acrodef{ids}[IDS]{intrusion detection system}

\begin{document}

%\setcounter{chapter}{2} % If you are doing your chapter as chapter one,
%\setcounter{section}{3} % comment these two lines out.

%\title{Deep Program Reidentification: A Graph Neural Network Solution}
\title{Attentional Heterogeneous Graph Neural Network:\\ Application to Program Reidentification}
\author{Shen Wang\thanks{Work done during an internship at NEC Labs America.} \thanks{University of Illinois at Chicago, \{swang224, psyu\}@uic.edu.}  \\
\and
Zhengzhang Chen\thanks{NEC Laboratories America, \{zchen, dingli, zhichun, ltang, jni, rhee, haifeng\}@nec-labs.com.} \thanks{Corresponding author.}
\and
Ding Li\footnotemark[2]
\and
Zhichun Li\footnotemark[2]
\and
Lu-An Tang\footnotemark[2]
\and
Jingchao Ni\footnotemark[2]
\and
Junghwan Rhee\footnotemark[2]
\and
Haifeng Chen\footnotemark[2]
\and
Philip S. Yu\footnotemark[1] %\footnotemark[1] \thanks{Institute of Data Science, Tsinghua University}
}
\date{}

\maketitle

% Copyright Statement
% When submitting your final paper to a SIAM proceedings, it is requested that you include 
% the appropriate copyright in the footer of the paper.  The copyright added should be 
% consistent with the copyright selected on the copyright form submitted with the paper.
% Please note that "20XX" should be changed to the year of the meeting.

% Default Copyright Statement
\fancyfoot[R]{\scriptsize{Copyright \textcopyright\ 2019 by SIAM\\
Unauthorized reproduction of this article is prohibited}}

% Depending on which copyright you agree to when you sign the copyright form, the copyright 
% can be changed to one of the following after commenting out the default copyright statement
% above.

%\fancyfoot[R]{\scriptsize{Copyright \textcopyright\ 20XX\\
%Copyright for this paper is retained by authors}}

%\fancyfoot[R]{\scriptsize{Copyright \textcopyright\ 20XX\\
%Copyright retained by principal author's organization}}

%\pagenumbering{arabic}
%\setcounter{page}{1}%Leave this line commented out.

\begin{abstract} 
%Program or process is the integral part of almost every IT/IoT system. Thus, program reidentification (\textit{i.e.}, checking the program's identity/ID) is often the very first step towards the safe and resilient system. Most, if not all, of the existing program reidentification methods, are solely based on the code-signing Public Key Infrastructure. However, recent research showed that compromised code-signing certificates have been associated with advanced threats like the Stuxnet worm. 
Program or process is an integral part of almost every IT/OT system. Can we trust the identity/ID (\textit{e.g.}, executable name) of the program? To avoid detection, malware may disguise itself using the ID of a legitimate program, and a system tool (\textit{e.g.}, PowerShell) used by the attackers may have the fake ID of another common software, which is less sensitive. However, existing intrusion detection techniques often overlook this critical \problem\ problem (\textit{i.e.}, checking the program's identity). 
%Malware can disguise itself as a legitimate program or hijack system processes to avoid detection.
In this paper, we propose an attentional heterogeneous graph neural network model (\ourss) to verify the program's identity based on its system behaviors. The key idea is to leverage the representation learning of the heterogeneous program behavior graph to guide the reidentification process. We formulate the program reidentification as a graph classification problem and develop an effective attentional heterogeneous graph embedding algorithm to solve it. Extensive experiments --- using real-world enterprise monitoring data and real attacks --- demonstrate the effectiveness of \ourss\ across multiple popular metrics and the robustness to the normal dynamic changes like program version upgrades.

\end{abstract}

\vspace{-6pt}
\section{Introduction}
% With computers and networked systems playing indispensable roles in almost every aspect of modern society such as industry, government, and economy, cyber security undoubtedly bears the utmost importance in preserving right social orders. 

% Security problem background.
% (1)Security fact.
% (2)Introduce traditional malicious program analysis 
% (3)Introduce the most common studied problem: malware detection, which is a binary classification problem.

% Introduce our newly formulated problem--program re-id.
% (1) drawback of malware detection: (a) too less malware sample (b) highly rely on the sandbox and takes long time anlysis (c) can not deal with unknown program (d) not robust to version upgrade
% (2) the reason to formulate the new problem
% (3) introduce the new problem and proposed approach
% (program plays the key role and it is the only system entity which can embedded hacker's malicious logic. Thus to detect the hacker's malicious behavior, we need to identify malicious program.
% Similar to person Re-ID problem, each program also has its own ID (normally its exename). It is necessary to check whether its ID matches its normal behavior or not to verify that it is really the one that it claimed itself to be. If it doesn't match, it is much more likely to have some problem. We should perform further analysis and matched with the database.)

% Introduce the application
% (1) Disguised program detection
% (2) Malicious program detection
% (3) Program update robustness

Modern enterprises often rely on \acf{ids}, either misuse-based or anomaly detection based, to protect their IT and OT systems. However, existing \ac{ids} techniques overlook one critical problem, which is the \problem: given a program, with a claimed ID (such as executable name), running in the system, can we confirm the program's identity and guarantee that this program is not a disguised malicious program, or a hijacked program with different behaviors, by comparing it to the normal program with the same ID? To make the attack processes stealthy and avoid detection, hackers often falsify the IDs of the tools they have used to bypass the \ac{ids}. For example, %a malware can disguise itself as a legitimate program
a malware may have the ID of a benign software and a system tool (\textit{e.g.}, PowerShell) used by the hackers may have the ID of another software which is less sensitive~\cite{liu2018towards}\nop{\cite{powershellinject}}. In fact, running programs with fake IDs is a strong signal of the system being compromised~\cite{gao2018saql}. Capturing the programs with fake IDs can help identifying very stealthy attacks and reduce the security risks in enterprise networks.

\nop{In order to avoid detection, malware can disguise
itself as a legitimate program or hijack system processes to reach
its goals. Commonly used signature-based Intrusion Detection
Systems (IDS) struggle to distinguish between these processes
and are thus only of limited use to detect such attacks. They also
have the shortcoming that they need to be updated frequently to
possess the latest malware definitions. This makes them inherently
prone to missing novel attack techniques. Misuse detection IDSs
however overcome this problem by maintaining a ground truth
of normal application behavior while reporting deviations as
anomalies.

There are many ways malware can avoid detection. One
such way is to disguise itself as a legitimate program or to
hijack system processes to reach its goals. This can start with
the visit of an infected website that exploits a vulnerability
in a browser plug-in [1], [2] and ranges over legitimate
downloads infected via a man-in-the-middle attack [3]. A
common technique to prevent plug-in based intrusion is sandboxing
of browser tabs [4]. Anti-virus software widely uses
signature based detection [5] to recognize infected executables
among incoming files. And several approaches to profile user
activities and detect deviation from normal behavior have been
established under the topic of continuous authentication [6],
[7], [8] as well as insider threat detection [9], [10].
A common detection evasion technique for malware is to
disguise itself as another process, either by hiding inside the
application’s binary or through direct impersonation
}

Existing techniques cannot be directly applied to address the problem of \problem. Digital code signing techniques, such as Public Key Infrastructure (PKI)~\cite{pki}, may help identify the certified authors of programs. However, many open-source programs that are widely used in enterprises may not contain valid signatures. Further, modern programs are evolving at a fast pace. Each version of the program has a unique signature. Handling the fast inflating set of signatures for programs is practically difficult. Malware detection or Anti-Virus\nop{~\cite{Zhang:2014:SAM:2660267.2660359}} may detect malware. Yet, hackers may also use common system tools to finish their attacks, such as malware free attacks~\cite{zimba2017malware}. Besides, a sophisticated malware can also bypass the anti-virus by hiding its malicious features, and a hacker can also hijack the memory of a benign program to perform malicious actions~\cite{arefi2018faros}. There are more sophisticated program watermarking techniques to identify a program~\cite{ren2014droidmarking}, but their computational costs are prohibitively high so that they cannot be widely applied to modern enterprise environments, which contain thousands of programs.

We observe that the system behaviors, such as file accessing, inter-process communications, and network communications, of a program have distinguishable patterns~\cite{tang2018node}. For example, every instance of excel.exe loads a fixed set of .DLL files while opening a spreadsheet file. If an EXCEL.EXE instance performs a rare operation, such as loading an unseen .DLL file or initiating another process, this excel.exe is very likely to be hijacked or even malware with the ID of EXCEL.EXE. Such patterns are stable during the evolution of the program, and the system behavioral events can be efficiently collected by system monitoring techniques~\cite{kernelaudit,etw}. 
Based on this observation, in this paper, we propose \ourss, an attentional \underline{deep} \underline{h}eterogeneous \underline{g}raph \underline{n}eural \underline{n}etwork based approach for program reidentificaion by modeling the system behaviors of the program. 
% system monitoring and surveillance data. 
%Specifically, to perform program identity verification efficiently and effectively, we propose a deep program identity verification approach via attentional multi-channel contextual graph encoder to generate the graph embedding of the program behavior graph. 

In particular, we design a compact graph modeling, the program behavior graph, to preserve all the useful information from massive system monitoring data and capture the interactions between different system entities. %, and construct a Heterogeneous Information Network (HIN). 
%This graph modeling effectively addresses three challenges (\textit{i.e.}, high volume, high dimension, and categorical) of the system event data and provides the foundation for further graph based program behavior analysis.
%Then, we develop a machine learning model to automatically learn the patterns of system behaviors of a program from the behavior graph. To the best of our knowledge, \ourss~is the very first approach that tries to address the \problem\ in an efficient and effective manner. 
%Learning the patterns from the behavior graphs, which is an HIN, is challenging. 
The constructed behavior graph is a heterogeneous graph \nop{~\cite{shi2017survey}}, which involves a hierarchy of different dependencies from simple to complex. %The behavior graph involves a hierarchy of different dependencies from simple to complex. 
Among all the dependencies, the complex ones are not exposed directly by the edges in the graph but can be inferred by a hierarchy of deep representation.
%\nop{\{Among all the dependencies, only simple dependencies  by edges are exposed, while the complex  ones are obscure but can be inferred by a hierarchy of deep representation.\}-----------------------}
%For example, a process accessing a file (P $\rightarrow$ F) is a simple dependency, two processes  accessing the same file (P $\rightarrow$ F $\leftarrow$ P) is a complex dependency. The complex dependency can not be modeled directly with shallow representation such as simple dependency, but a hierarchy of deep representation, from simple dependency (P $\rightarrow$ F) to (P $\rightarrow$ F $\leftarrow$ P). 
%\nop{To handle the complex and obscure dependencies, we need a deep graph structure learning model to capture the hierarchical dependency. However, recent deep graph structure learning methods~\cite{kipf2016semi,defferrard2016convolutional,bruna2013spectral,hamilton2017inductive,velickovic2017graph} only works for homogeneous undirected graph and can not capture the heterogeneity of \textbf{HIN}.}
%To address this challenge, an attentional multi-channel contextual graph encoder is proposed to generate the graph embedding of the program behavior graph. 
%To make the heterogeneous graph learnable via a deep representation learning, **************************
%To simplify the learning curve of graph neural network in heterogeneous graph
To capture the hierarchical dependencies, we first propose a multi-channel transformation module to transform the heterogeneous graph into the multi-channel graph guided by the meta-paths. 
%To leverage the information of each entity, we extract the knowledge-free graph features. 
After the multi-channel transformation, we feed the resulted graph and its corresponding entity features into a graph neural network for graph embedding. 
%However, each homogeneous graph is directed, adapting existing graph neural networks directly fails to capture the directed dependency. To address this challenge, 
We propose a contextual graph encoder and stack it layer by layer to learn the hierarchical graph embedding via propagating the contextual information. Noticing the different importance of the different channels, channel-aware attention is further developed to align the multi-channel graph embeddings.   
We conduct an extensive set of experiments on real-world enterprise monitoring data to evaluate the performance of our approach. The results demonstrate the effectiveness \nop{and efficiency} of our proposed algorithm. We also apply \ourss\ to real enterprise security systems. The results show our method is effective in identifying the disguised signed programs and robust to the normal dynamic changes like program version upgrade.

In summary, the contributions of this paper are:%this paper makes the following contributions:
\begin{itemize}
\vspace{-8pt}
\item %\textbf{Meaningful Newly Formulated Problem Setting} 
We identify the important problem of program reidentification in intrusion detection, which is often overlooked by the existing intrusion detection systems;
\vspace{-8pt}
\item %\textbf{Novel Graph Modeling of Program Behavior}
We propose a heterogeneous graph model to capture the interactions between different system entities from large-scale system surveillance data;
\vspace{-8pt}
\item %\textbf{Novel Multi-Channel Learning Framework for HIN}
We develop a multi-channel transformation to transform a heterogeneous information network into a multi-channel graph; 
\vspace{-8pt}
\item %\textbf{Effective and Efficient Representation Learning from Directed Graph}
We propose a heterogeneous graph neural network based approach to learn the graph embedding via propagating the contextual information;
\vspace{-8pt}
\item %\textbf{Effective Channel-Aware Attention Mechanism}
We propose a channel-aware attention mechanism to learn the representation from different channels jointly; 
\vspace{-8pt}
\item %\textbf{Effective Channel-Aware Attention Mechanism}
Our empirical studies on real enterprise monitoring data demonstrate the effectiveness of our method.
\end{itemize}

\vspace{-15pt}
\section{Preliminaries and Problem Statement}
\label{sec:pre}
In this section, we first present the preliminaries, then define the machine learning problem that we are concerned with \ours. %For the following sections, we assume that the operation system is Windows System, for simplicity of explanation. 
%\subsection{Notations}

\noindent{\bf{System Entity}}
%A system entity is the basic element of the every operating system. 
There are three main types of system entities in an operating system \cite{CaoCCTLL18,LinCCT0CL18}: \textit{processes}, \textit{files}, and \textit{Internet sockets} (INETSockets). And each entity is associated with a set of categorical attributes. In this paper, we use ``program'' and ``process'' interchangeably whenever there is no ambiguity. %For the file entity, we use its extension name as the identifier.

\noindent{\bf{System Event}}
A system event is an interaction between a pair of system entities. Formally, a system event $e=(v_s, v_d, t)$ represents a source entity $v_s$, destination entity $v_d$, and their interaction happens at time stamp $t$. There are multiple types of system events, due to the existence of different types of entities. We consider three different types of system events, including: (1) a process forking another process (P-P), (2) a process accessing a file (P-F), and (3) a process connecting to an INETSocket (P-I). %These three types of system events model the causal dependency from a process to another system entity. Each system event is also associated with a set of attributes, including the information of the involved system entities and their dependency.

\noindent{\bf{Problem Statement}}
%Formally, we define our problem as followed. 
%Given a number of system snapshots $S$ of a program with claimed , we check whether the process name matches the normal behavior of claimed program and suggest the top-k program.
%Given a system snapshots $S$ of a program, we check whether it belongs to the program in the database and suggest the top-k programs. 
Given a target program with corresponding event data during a time window $U = \{e_1,e_2,...\}$ and a claimed name/ID, we check whether it belongs to the claimed name/ID. If it matches the behavior pattern of the claimed name/ID, the predicted label should be $+1$; otherwise it should be $-1$. More formally,  %we formulate it as a binary classification problem: 
given event data $U$ of a program, a mapping function $f$ is used to map $U$ to a binary label $Y\in \{+1,-1\}$, such that $f:U\rightarrow \{+1,-1\}$.%Due to the unique characteristic of the program event data, we propose attentional multi-channel graph neural network to model this mapping function.

\vspace{-6pt}
\section{Algorithm}

\begin{figure*}[htb]
\centering
\vspace{-15pt}
\includegraphics[width =1\linewidth]{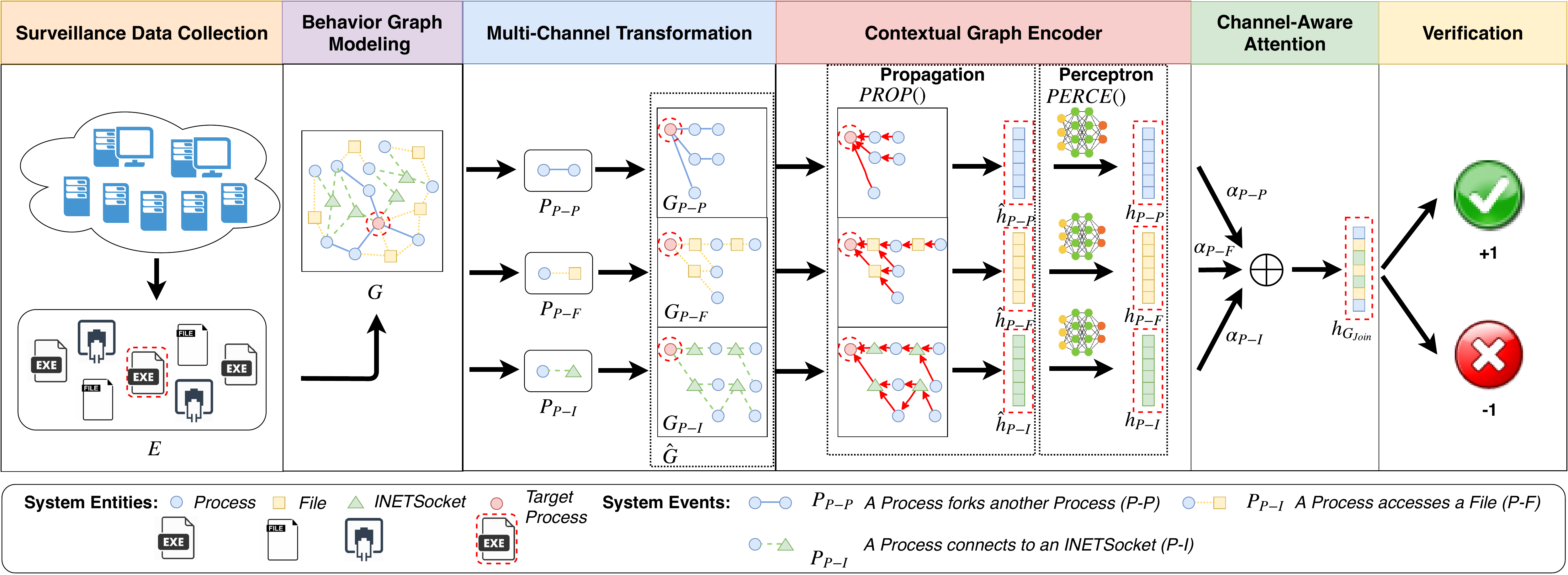}
\caption{System architecture of deep program reidentification.}
\label{fig:Architecture}
\vspace{-10pt}
\end{figure*}
%\myred{TODO: The method is too long. will cut at least 0.5-1 page, especially the behavior graph modeling subsection}
\subsection{Overview}
In this section, we introduce \ourss, a graph neural network based approach to verify the system program in a data-driven manner. The framework of \ourss\ (as shown in Figure \ref{fig:Architecture}), consists of the following six components:
\begin{itemize}
\vspace{-2pt}
\item \textbf{Surveillance Data Collection}.
%The low-level system events are collected from a enterprise network. 
This module collects three different types of system events (see Section \ref{sec:pre} for definition) from IT/OT systems; %including a process folks another process (P$\rightarrow$P), a process access a file (P$\rightarrow$F), and a process connects to INETSocket (P$\rightarrow$I) across three types of system entities.
%Three type of system entities are considered, including process, file and INETSocket.
\vspace{-5pt}
\item \textbf{Behavior Graph Modeling}.
Based on the collected program event data, a compact heterogeneous program behavior graph is constructed for the target process to capture the complex interactions and eliminate the data redundancies;
%Since the raw attribute of each entity are extreme high dimensional and categorical, which are difficult to use without any expert knowledge, we extract the knowledge-free feature for each entity.  
\vspace{-8pt}
\item \textbf{Multi-Channel Transformation}.
This module transforms the generated heterogeneous behavior graph to a multi-channel graph with each channel modeling one type of meta-path relationship; 
%The knowledge-free feature are extracted for each entity within each channel graph.  
\vspace{-8pt}
\item \textbf{Contextual Graph Encoder}.
Based on the generated multi-channel graph, we propose an effective graph neural network with propagation layer and perceptron layer to extract the graph embedding; %from the target process context. 
%the directed graph embedding. Benefit from flexible context propagation, we can capture the directed causal relation.
\vspace{-8pt}
\item \textbf{Channel-Aware Attention}.
In this module, an attention mechanism is proposed to align the graph embeddings extracted from different channels and generate the attentional joint embedding;

\vspace{-8pt}
\item \textbf{Program Reidentification}
After the joint embedding being extracted, the resulted  low-dimensional vector is fed to a classification layer to generate the prediction. %Specifically, we use logistic regression at classification layer.
\vspace{-8pt}
\end{itemize}

\subsection{Behavior Graph Modeling}
\label{subsec:graph}
%When a program is executed in the operation system, we can produce system events data by monitoring the behavior of the program. By observing and analyzing the system event data, we found that they has four properties: 
Real information networks like the enterprise networks often generate a large volume of system behavioral data with rich information on program/process level events. We utilize a system monitoring tool (\textit{i.e.}, Windows ETW~\cite{etw})\nop{ and Linux audit framework~\cite{kernelaudit}} to collect the program behavioral event data. The data has four properties:
(1) High Volume. The system event data collected from a single computer system 
%by monitoring the process interactions 
in one day can easily reach 20 GB and the number of events relates to one specific program can easily reach thousands. It is prohibitively expensive to perform the analytic task on such massive data in terms of both time and space; 
(2) High Dimensionality. Each system event is associated with hundreds of attributes, including information of involved system entities and their relationships, which causes the curse of dimensionality \cite{verleysen2005curse}; 
(3) Categorical. 
All available attributes are categorical, and each attribute has hundreds of categories. 
%For example, in a single computer system, the number of the filename and the number of IP address accessed within one week can easily reach 10k. 
Analysis of categorical data is very challenging due to the lack of intrinsic proximity measure \cite{zhang2015categorical}; 
(4) Redundant. 
The attributes are redundant. 
Most of the attributes of system entities are the same. Repeatedly saving these attributes results in a significant redundancy. The events are also redundant. The events with the same system entity and relationship are usually repeatedly stored, which also causes a notable redundancy.
These challenges motivate our idea of graph modeling.  
%Therefore, we devise a compact graph representation of the system event data.
%By observing and analyzing the system event data, we found that they are usually redundant. The redundancy comes from several sources. 
%Firstly, the attributes are redundant. Most of the attributes of system entities are same. Repeatedly saving these attributes results a significant redundancy.   
%Furthermore, the events are redundant. The event with same system entity and relationship are usually repeatedly stored, which also causes a notable redundancy. These redundancy motivates our idea of graph modeling. 

We construct the compact graph to model the program behaviors. Formally, given the program event data $U$ across many machines within a time window (\textit{e.g.}, 1 day), a heterogeneous graph $G_{Behavior} = (V_{Behavior},E_{Behavior})$ is constructed for the target program. $V_{Behavior}$ denotes a set of nodes, with each representing an entity of three types: process (P), file (F), and INETSocket (I). Namely, $V_{Behavior} = P \cup F \cup I$. %The number of entity is $n_v$.
$E_{Behavior}$ denotes a set of edges $(v_s,v_d, r)$ between the source entity $v_s$ and destination entity $v_d$ with relation $r$. We consider three types of relations, including: (1) a process forking another process (P$\rightarrow$P), (2) a process accessing a file (P$\rightarrow$F), and (3) a process connecting to an Internet socket (P$\rightarrow$I). Each graph is associated with an adjacency matrix $A$. 
%Since the constructed graph is directed, $(v_s,v_d, r) \neq (v_d,v_s, r)$. 
With the help of the behavior graph modeling, we reduce the data redundancy significantly but keep important information.
\vspace{-8pt}
\subsection{Attentional Heterogeneous Graph Neural Networks}
\label{subsec:att}
After the graph modeling, a heterogeneous program behavior graph is constructed. There exist a hierarchy of different dependencies from simple to complex. Among all the dependencies, the complex ones are not exposed directly by the edge in the graph but can be inferred by a hierarchy of deep representation. For example, a process accessing a file (P$\rightarrow$F) is a simple dependency, two processes  accessing the same file (P$\rightarrow$F$\leftarrow$P) is a complex dependency. %The complex dependency cannot be modeled directly with shallow representation such as a simple dependency, but can be inferred from a hierarchy of deep representation, from simple dependency (P$\rightarrow$F) to (P$\rightarrow$F$\leftarrow$P). 
To infer the relationship of two processes accessing the same file, we can check whether they share the same context. Due to the complex and obscure dependency, it requires a deep graph structure learning model to capture the hierarchical dependency. However, recent deep graph neural network methods \cite{kipf2016semi,defferrard2016convolutional,hamilton2017inductive,velickovic2017graph} only focus on the homogeneous graph and can not capture the heterogeneity within the heterogeneous graph.
%There are multiple type of entities and multiple type of connections. The diversity among different entities and edges are large. 
To learn the hierarchical dependency representation (hierarchical graph embedding) from heterogeneous graph, we propose an attentional heterogeneous graph neural network, which consists of multi-channel transformation, input layer, contextual graph encoder, and channel-aware attention. %Here, we introduce these components one by one and build the model step by step. 

\noindent{\bf{Multi-channel Transformation}} %Due to the heterogeneity intrinsic of nodes and edges in a hterogeneous graph, the diversities between different dependencies vary dramatically, which significantly increased the difficulty of applying graph neural network.
Due to the heterogeneity intrinsic of entities (nodes) and dependencies (edges) in a heterogeneous graph, the diversities between different dependencies vary dramatically, which significantly increase the difficulty of applying graph neural network.
%To make the heterogeneous graph learnable via a deep representation learning, 
To address this challenge, we design a multi-channel transformation module to transform the heterogeneous graph to a multi-channel graph guided by the meta-paths.
A meta-path \cite{sun2011pathsim} is a path that connects entity types via a sequence of relations over a heterogeneous network. For example, in a computer system, a meta-path can be the system events (P$\rightarrow$P, P$\rightarrow$F, and P$\rightarrow$I), with each one defining a unique relationship between two entities.

\nop{modification2--Notice that, there are many ways to construct meta paths: At the beginning, meta paths were constructed manually by the experts; Afterwards, several efficient methods were presented
to construct meta paths automatically, such as \cite{chen2017task}. In this paper, We follow it to construct meta paths.}

%and can be used as the link in a specific channel in the multi-channel graph. 
\nop{
(1) relationship of a process forking another process (P $\rightarrow$ P),
(2) relationship of a process accessing a file (P $\rightarrow$ F),
(3) relationship of a process opening an Internet socket (P $\rightarrow$ I),
(4) relationship of two processes accessing the same file (P $\rightarrow$ F $\leftarrow$ P), 
(5) relationship of two processes opening the same Internet socket (P $\rightarrow$ I $\leftarrow$ P).} 
%Specifically, we model the inbound neighborhoods and outbound neighborhoods separately in different channels for each type of directed connection.    
The multi-channel graph is a graph with each channel constructed via a certain type of meta-path. Formally, given a heterogeneous graph $\mathcal{G}$ with a set of meta-paths $M=\{M_1, ... ,M_{|C|}\}$, the transformed multi-channel network $\hat{G}$ is defined as follows:
\vspace{-8pt}
\begin{equation}
\hat{G}=\{G_i | G_i=(V_i,E_i, A_i) , i=1,2,...,|C|) \}
\end{equation}
%\vspace{-2pt}
where $E_i$ denotes the homogeneous links between the entities in $V_i$, which are connected through the meta-path $M_i$. Each channel graph $G_i$ is associated with an adjacency matrix $A_i$. $|C|$ indicates the number of meta-paths. Notice that, the potential meta-paths induced from the heterogeneous network $G_{Behavior}$ can be infinite, but not everyone is relevant and useful for the specific task of interest. Fortunately, there are some algorithms \cite{chen2017task} proposed recently for automatically selecting the meta-paths for particular tasks.

\noindent{\bf{Input Layer}}
After the multi-channel transformation, an input layer is applied to construct the feature vector for each entity. Specifically, two types of features are constructed: connectivity features and statistic features. Since these features are constructed without any expert knowledge, they are knowledge free. 

Connectivity features are constructed to describe the pairwise proximity between entities. The connectivity feature of entity $v$ is defined as $X^{con}_{v}=\{e_{v, 1}...,e_{v, |V|}\}$, which denotes the first-order proximity between $v$ and other entities. Graph statistical feature of an entity $v$ is generated based on the graph theory. It can be defined as $X^{stat}_{v}=\{X_v^{s1},X_v^{s2},X_v^{s3},X_v^{s4}\}$, including four graph measurements: degree centrality, closeness centrality, betweenness centrality, and clustering coefficient.

\noindent{\bf{Contextual Graph Encoder}}\footnote{In this subsection, we remove the subscripts of channel indicator for all the graph related symbols for simplification.}
In this step, we feed the multi-channel graph and the corresponding entity features to a graph neural network for graph embedding. To learn the hierarchical graph representation, we propose contextual graph encoder (CGE) and stack a deep graph neural work.
%In this section we introduce the proposed novel GNNs Model, named Contextual Graph Encoder (CGE) in order to learn deep graph representation from each channel of homogeneous directed graph.
Given the one-channel graph $G = (V, E, A)$ with each $V$ associated with a corresponding feature $X$, our target is to learn an encoding function $f_{G}:G \rightarrow h_{G}$, that encodes the graph to a low dimensional vector (graph embedding), where $h_{G}$ denotes the generated graph embedding vector. Specifically, we use the representation of the target entity $h_{v_{t}}$ (target contextual entity embedding), which encodes both contextual structure and its features by a mapping function $f_{V}:V \rightarrow h_{V}$, as the representation of the graph. In this way, we reduce the graph embedding problem into the target entity embedding. Formally, our graph encoder can be defined as follows:
\vspace{-5pt}
\begin{equation}
h_{G} = h_{v_{t}} = f_{V}(V_t)
\end{equation}
The basic contextual graph encoder consists of a propagation layer $PROP()$ and a perceptron layer $PERCE()$. The propagating layer propagates the information from the context of the target entity. We define the propagation layer $PROP()$ as follows:
\vspace{-5pt}
\begin{equation}
\hat{h}^l=PROP(h^l)=P^lh^l
\end{equation}
where $l$ denotes the layer number, $P^l$ denotes the propagation matrix at layer $l$, and 
$\hat{h}^{l}$ indicates the output of $PROP()$ at layer $l$. The first layer $h^0=X$ takes the features of each entity. 
The propagation layer propagates information to the target entity within a region, named graph receptive field $\mathcal{F}$. The graph receptive field $\mathcal{F}$ usually consists of all $L$-hop contexts of target entities. The propagation operation encodes both the graph structure and the entity features, which is similar to performing a graph convolution operation \cite{kipf2016semi}. After the propagation is performed, a perceptron layer $PERCE()$ is applied to the propagated representation of the entities, such that:
\vspace{-5pt}
\begin{equation}
h^{l+1}=PERCE(\hat{h}^l)=\sigma(\hat{h}^lW^l)
\end{equation}
where $W^l$ is the shared trainable weight matrix for all the entities at layer $l$ and $\sigma()$ is a nonlinear gating function. Benefit from weight sharing, it is both statistically and computationally efficient compared with traditional entity embedding. Weight sharing can act as powerful regularization to preserve the invariant property in the graph, and the number of parameters is significantly reduced.

We design our propagation layer by performing the random walk on the graph via a diffusion process characterized by a specific probability $q \in [0,1]$ and a state transition matrix $D^{-1}A$. Here, $D$ denotes the degree matrix of the adjacency matrix $A$, such that $D=diag(A \textbf{1})$. \textbf{1} is a all one vector. After some transitions, such Markov process converges to a stationary distribution $P\in\mathcal{R}^{N \times N}$ with $i$th row indicates the likelihood of diffusion from the entity, hence the proximity of that entity. This stationary distribution of the diffusion process is proven to have a closed form solution \cite{atwood2016diffusion}. When considering the 1-step truncation of the diffusion process, the propagation layer is defined as follows:
\vspace{-5pt}
\begin{equation}
\hat{h}^l=PROP(h^l)=D^{-1}Ah^l=\sum_{u\in N(v_t)}P_{uv_t}h^l
\end{equation}
The receptive field is defined as $\mathcal{F}=\{N(v_t)\}$, with $N(v_t)$ denoting all the contexts of target entity $v_t$. The propagation layer computes the weighted sum of the contexts' current representation. We set $P^0=P^1=P^2=D^{-1}A$ and build a three-layer CGE as follows:
%This is equivalent to replace the original propagation matrix $P$ with a sparser unbiased estimator $\hat{P}$,  $E$ is the sparser unbiased estimator such that $E (P) = \hat{P}^l$, where $\hat{P}^l_{uv_t}=\frac{n(v_t)}{|n^l(v_t)|} P_{uv_t}$ if $u \in n^l(v_t)$. we propagating over
\vspace{-5pt}
\begin{align}
h^1 = \text{ReLU}((P^0 X) W^0)\\
h^2 = \text{ReLU}((P^1 h^1) W^1)\\
h^3 = \text{ReLU}((P^2 h^2) W^2)\\
h_{G} = h_{v_{t}} = h^3
\label{eq:sp}
\vspace{-5pt}
\end{align}
where $\text{RELU}()$ is a element-wise rectified linear activation. We perform the CGE to each channel graph and generate corresponding graph representation respectively.

\noindent{\bf{Channel-Aware Attention}}
Going through the CGE, the graph embeddings for each channel graph are extracted. However, different channels should not be considered equally. For example, Ransomware is usually very active in accessing the files, but it barely forks another process, or opens an internet socket, while the VPN is generally very active in opening the internet socket, but it barely accesses a file or forks another process. Therefore, we need to treat different channels differently. Here, we propose a channel-aware attention, an attention mechanism, to align the graph embeddings from different channels and generate the joint embedding. Specifically, we learn the attention weights for different channels automatically.

Formally, given the corresponding graph embedding $h_{G_i}$ for each channel $i=1,2,...,|C|$, we define the attention weight as follows:
\vspace{-5pt}
\begin{equation}
\alpha_{i}=\frac{\exp (\sigma(a[W_a h_{G_i}||W_a h_{G_k}]))}{\sum_{k'\in |C|} \exp (\sigma (a[W_a h_{G_i}||W_a h_{G_{k'}}]))}
\label{eq:att}
\end{equation}
where $h_{G_{i}}$ is graph embedding of the target channel, $h_{G_{k}}$ denotes the representation of the other channels. $a$ denotes a trainable attention vector, $W_{a}$ denotes a trainable weight mapping the input features to the hidden space, $||$ denotes the concatenation operation, and $\sigma$ denotes the nonlinear gating function. We formulate a feed-forward neural network that is used to compute the correlation between one channel and other channels. This correlation is normalized by a Softmax function. Let $ATT(h_{G_{i}})$ represent Eq.(\ref{eq:att}). The joint representation of each channel can be represented as follows:
\vspace{-8pt}
\begin{equation}
h_{G_{Join}}=\sum_{i=1}^{|C|} ATT(h_{G_i}) h_{G_i}
\end{equation}
The channel-aware attention allows us to better infer the importance of different channels by leveraging their correlations and learn a channel-aware representation. 
%In this way, we can mode the different type of program system level behavior differently. 
\vspace{-5pt}
\subsection{Program Reidentification}
After the joint graph embedding is generated from program event data, a binary classifier is used to verify whether the program matches its claimed name. Specifically, we trained the binary classifier for each known program. Our framework takes a claimed program event data as the input and generates the corresponding prediction. The final output is +1 or -1, indicating the identified or unidentified prediction result. 

To train a verification model for a particular program, we collect a set of program events $\mathcal{U}=\{U_1, U_2, ..., U_m\}$ including events belong to that program and the ones do not belong to that program. Their corresponding labels are $Y=\{y_1,y_2,...,y_m\}$ with $y_i \in \{+1, -1\}$. If the event data belongs to the claimed program, its ground truth label  $y_i=+1$, otherwise its ground truth label $y_i=-1$.
Our target is to design an end-to-end binary classifier. 
Specifically, we propose to use logistic regression classifier, and the objective function is defined as follows:
\vspace{-5pt}
\begin{equation}
l=\frac{1}{m}\sum_{i=1}^{m} [y_i\log \hat{y}_i + (1-y_i)\log (1-\hat{y}_i)]
\end{equation}
where $\hat{y}_i$ denotes the predicted label. We optimize the above objective with Adam optimizer. The gradients of the parameters are calculated recursively according to the graph topology. Since our approach is end to end, we directly optimize the reidentification objectives. Once the model achieves a good performance (\textit{e.g.}, using accuracy (ACC) as the measure), the training process terminates, and the trained model is suitable for program reidentification.
\vspace{-6pt}
\section{Experiments}
%In this section, we evaluate the proposed method on real system surveillance data collected from a real enterprise network. We first introduce the datasets used on the experiments and the experimental setting. Then we show the effectiveness of the proposed method. We first conduct the synthetic experiment. Then we apply it in  real-world tasks: disguised program detection. Finally, we show the robustness of our approach to the program version upgrade.  
\subsection{Data}
\label{subsec:data}
In the experiments, we evaluate the proposed method on the real-world system monitoring data. The data is collected from a real enterprise network composed of $87$ machines, in a time span of $20$ consecutive weeks. The sheer size of the data set is around 3 Terabytes. We consider three different types of system events as defined in Section~\ref{sec:pre}. Each entity is associated with a set of attributes, and each process has an executable name as its identifier/ID. In total, there are about $300$ million event records, with about $2,000$ processes, $600,000$ files, and $18,000$ Internet sockets. Based on the system event data, we construct the heterogeneous behavior graph (see Section~\ref{subsec:graph}) per program per day. For each entity, we construct three different types of features according to Section~\ref{subsec:att}:  \textit{fea-1}: connectivity feature,  \textit{fea-2} statistics feature, and \textit{fea-3}: the combination of \textit{fea-1} and \textit{fea-2}. %For statistic feature, we use four graph measurements, including degree centrality, closeness centrality, betweenness centrality and clustering coefficient .

\nop{
The statistic of the dataset is summarized in Table \ref{tab:dataset}.

\begin{table}[]
\centering
\begin{tabular}{|l|l|l|l|}
\hline
Events       & P-P & P-F & P-I \\ \hline
\#Agent      & 87      & 86       & 84     \\ \hline
\#SEntity    & 1902    & 1935     & 142     \\ \hline
\#DEntity    & 1904    & 69813    & 16466   \\ \hline
\#Edge       & 8    &     &     \\ \hline
\#Avg.Degree &     &     &     \\ \hline
%\#Feature    &     &     &     \\ \hline
\end{tabular}
\caption{The statistics for each type of events}
\label{tab:dataset}
\end{table}
}
\nop{
\myred{check the of Ransomeware}
}
\vspace{-5pt}
\subsection{Experiment Setup}
\nop{
\subsubsection{Feature Engineering}
For each program entity, we can extract two both connectivity feature and graph statistics feature. In the experiment, we construct three type of features: \textit{fea-1}: connectivity features; \textit{fea-2}: graph statistics features ; \textit{fea-3}: combinations of connectivity features and graph statistics features.
}
\subsubsection{Baselines}
\vspace{-5pt}
\nop{To evaluate the performance of \ours,}We compare the proposed method \ourss\ with the following typical and state-of-art classification models:
\vspace{-5pt}
\begin{itemize}
\item \textit{LR} and \textit{SVM}: LR and SVM represent the Logistic Regression and Linear Support Vector Machine, respectively. They are two typical classification methods. The raw features extracted from each process are used as the input, including the connectivity features and the graph statistics feature. The LR and SVM are implemented using sciket-learn.
\vspace{-5pt}
\item \textit{XGB}: XGB represents the gradient boosting. It is a decision tree based classification model implemented using XGBoost \cite{chen2016xgboost}. It is the state-of-art linear classification model for most of the tasks. We set maximum $500$ trees with the learning rate equals to $0.1$.
\vspace{-5pt}
\item \textit{MLP}: MLP represents the Multi-layer Perceptron, which is a deep neural network based classification model with multiple non-linear layers between the input layer and the output layer. It is a special case of our contextual graph encoder if we define the propagation layer as an identity matrix. 
\end{itemize}

\nop{
Since the proposed approach is based on graph neural network. We compare our method with following graph neural networks methods. 

\begin{itemize}
\nop{
\item \myred{should we compare with traditional node embedding, that's a new work}
}
\item \textit{GCN}:GCN \cite{kipf2016semi} is graph convolutional based semi-supervised classification algorithm. It is a special case of our contextual graph encoder. It consider all the 1-hop neighbourhoods of target entity as the receptive field in each graph encoder and extract the deeper relational feature through stacked multiple convolutional layer. It focus on homogeneous network.
% \item High Uncertainty: There are more than thousand of program in a machine, which causes high uncertainty 
\item \textit{GraphSage}:GraphSage \cite{hamilton2017inductive} is another convolutional-like graph neural network. It is also a special case of our contextual graph encoder. Its receptive field is constructed based on the neighborhood sampled using uniform distribution. It also focus on homogeneous network.

\end{itemize}
}
Since our approach \ourss\ consists of six components, we consider different variants as well. 
\nop{
of our approach, including \textit{\ourss-pp}, \textit{\ourss-pf}, \textit{\ourss-pi}, \textit{\ourss-con}. \textit{\ourss-pp}, \textit{\ourss-pf} and \textit{\ourss-pf} denotes our method only consider single channel. \textit{\ourss-pp} denotes our method consider all channels, but joint different channel with direct concatenation. Our method is implemented with Tensorflow.
}
% \begin{itemize}
% \item \textbf{\ourss-pp}: It is our proposed method only consider P-P channel.
% \item \textbf{\ourss-pf}: It is our proposed method only consider P-F channel.
% \item \textbf{\ourss-pi}: It is our proposed method only consider P-I channel.
% \item \textbf{\ourss-all}: It is our proposed method consider all channels but joint different channel with direct concatenation. 
% \item \textbf{\ourss}: This our proposed method, which consider all channels and joint them together with attentional weight. 
% \end{itemize}
\vspace{-5pt}
\subsubsection{Evaluation Metrics}
Similar to \cite{LinCCT0CL18,CaoCCTLL18}\nop{Akoglu:2012}, we evaluate the performance of different methods using a variety of measures, including accuracy (ACC), F-1 score, AUC score, precision, and recall.
% Table 1.

% \begin{table}[]
% \centering
% \scalebox{0.8}{
% \begin{tabular}{|l|l|}
% \hline
% Metrics   & Description                                  \\ \hline
% TP        & \# of target program correctly predicted     \\ \hline
% TN        & \# of other program correctly predicted      \\ \hline
% FP        & \# of false prediction as the target program \\ \hline
% FN        & \# of false prediction as other program      \\ \hline
% %ACC       & (TP+TN)/(TP+TN+FP+FN)                        \\ \hline
% Precision & TP/(TP+FP)                                   \\ \hline
% Recall    & TP/(TP+FN)                                   \\ \hline
% F1        & 2*Precision * Recall/(Precision + Recall)    \\ \hline
% AUC       &                                              \\ \hline
% %TPR       & True positive rate, FP/(TN+FP)               \\ \hline
% %FPR       & False positive rate, FN/(TP+FN)              \\ \hline
% \end{tabular}}
% \caption{Performance metrics}
% \label{tab:metric}
% \end{table}
\vspace{-5pt}
\subsection{Synthetic Experiments}
To evaluate the proposed method in a more controlled setting\nop{for assessing algorithm performance}, we conduct three sets of synthetic experiments on $500$ most active programs as follows:
(1) We evaluate the effectiveness of multi-channel transformation and channel-aware attention modules;%We compare with configuration with single meta-path and their combinations.
\nop{
(2) we evaluate the effectiveness of our proposed contextual neural graph encoder module and compare it with its different variations.
}
(2) We evaluate the performance of our method on normal program reidentification;
(3) We perform the parameter sensitivity analysis.
%We call them synthetic experiments, since all the programs we used in training and testing are normal program with normal behavior. 
%Based on the collected system event data, we first generate one behavior graph per program per time window (\textit{i.e.}, one day). Then, for the target program, we randomly sample $5,00$ behavior graphs generated from different time windows. belonging to that program and another $5,00$ belonging to the rest of programs to have balanced configuration. 
We perform 5-fold cross-validation on each program and report the average testing results of all the programs for each evaluation metric. We conduct a grid search on the parameter of each method to identify the parameter setting that yields the best result, which is done using cross-validation. In particular, we set the dimension of the hidden layer to $500$.
%\vspace{-15pt}
\subsubsection{Evaluation of Different Meta-paths}
\begin{table}[htb]
\vspace{-15pt}
\centering
\begin{tabular}{|l|lll|}
\hline
\multicolumn{1}{|c|}{\multirow{2}{*}{Meta-Path}} & \multicolumn{3}{l|}{Evaluation Criteria} \\ \cline{2-4} 
\multicolumn{1}{|c|}{}                           & ACC          & F-1         & AUC         \\ \hline
$\ourss_{pp}$                                              &    0.838          &    0.864         &     0.843       \\
$\ourss_{pf}$                                              &    0.821          &    0.855         &     0.838       \\
$\ourss_{pi}$                                              &    0.579          &    0.635         &     0.592         \\
$\ourss_{con}$                                          &       0.876       &     0.901        &     0.890        \\
$\ourss_{att}$                                          &       \textbf{0.905}       &     \textbf{0.929}        &     \textbf{0.908}        \\ \hline
\end{tabular}
\caption{Reidentification results of different meta-paths.}
\label{tab:meta-path}
\vspace{-10pt}
\end{table}
In this experiment, 
we evaluate the performance of different kinds of meta-paths and their combinations. Giving each kind of meta-path, such as P$\rightarrow$P, P$\rightarrow$F, or P$\rightarrow$I, corresponding to one type of system event, we construct the multi-channel graph with entity features and then feed them to the contextual graph encoder to generate the graph embedding. After the graph embedding obtained, we feed it to logistic regression to train the classification model for program reidentification. We denote these baselines as $\ourss_{pp}$, $\ourss_{pf}$, and $\ourss_{pi}$. 
%When considering the combination of different meta-paths, we combine the graph embedding generated from different meta-paths. 
%For the combination of different meta-paths, 
To effectively evaluate our attention module,
we consider both direct concatenation and our proposed attentional version, which are denoted as $\ourss_{con}$ and $\ourss_{att}$.
%The experimental results are shown in Table \ref{tab:meta-path}. 
From Table \ref{tab:meta-path}, %we can see that different meta-paths and their combinations show different performance since each of them represents a specific type of dependency. We 
we can observe that: (1) The $\ourss_{att}$ outperforms \ourss\ with any single type of meta-path and their simple combination; (2) The combinations of multiple meta-paths perform better than any single meta-path, since they cover multiple types of dependency and provide complementary information; (3) The one with channel-aware attention performs better than simple concatenation, since the importance of different meta-paths vary for different programs. %Some programs may have active P$\rightarrow$P events, while the others may not. 
Channel-aware attention captures that difference and computes a weighted combination of different meta-paths. %(4) The P$\rightarrow$P and P$\rightarrow$F type meta-paths outperform the P$\rightarrow$I type meta-path, since most of the programs have active P$\rightarrow$P and P$\rightarrow$F events, while only a small part of programs have active P$\rightarrow$I events.

\vspace{-5pt}
\subsubsection{Evaluation on Normal Program Reidentification}
\begin{table*}[htb]
\small
\centering
\vspace{-20pt}
\begin{tabular}{|l|l|lllll}
\hline
\multicolumn{1}{|c|}{\multirow{2}{*}{Method}} & \multirow{2}{*}{Settings} & \multicolumn{5}{c|}{Evaluation Criteria} \\ \cline{3-7} 
\multicolumn{1}{|c|}{} &  & ACC & F-1 & AUC & Precision & \multicolumn{1}{l|}{Recall} \\ \hline
\multirow{3}{*}{LR} & \textit{fea-1} &0.693  & 0.755 & 0.699  &0.632  & \multicolumn{1}{l|}{0.948} \\
 & \textit{fea-2} & 0.705 & 0.770 & 0.703  & 0.655 & \multicolumn{1}{l|}{0.950} \\
 & \textit{fea-3} & 0.724 & 0.772 & 0.727 & 0.675 & \multicolumn{1}{l|}{0.948} \\ \hline
\multirow{3}{*}{SVM} & \textit{fea-1} &0.502  &0.662  &0.502  &0.505 & \multicolumn{1}{l|}{0.970} \\
 & \textit{fea-2} & 0.795 & 0.778 & 0.725 & 0.701 & \multicolumn{1}{l|}{0.935} \\
 & \textit{fea-3} & 0.504 & 0.652 & 0.504 & 0.505 & \multicolumn{1}{l|}{\textbf{0.975}} \\ \hline
\multirow{3}{*}{XGB} & \textit{fea-1} & 0.775  & 0.802 & 0.776 & 0.732 &  \multicolumn{1}{l|}{0.930} \\
 & \textit{fea-2} & 0.833 & 0.860 & 0.846 & 0.821 & \multicolumn{1}{l|}{0.936} \\
 & \textit{fea-3} & 0.855 & 0.866 & 0.856 & 0.827 & \multicolumn{1}{l|}{0.937} \\ \hline
\multirow{3}{*}{$MLP_{shallow}$} & \textit{fea-1} &0.633  & 0.745 & 0.643 & 0.626 & \multicolumn{1}{l|}{0.938} \\
 & \textit{fea-2} & 0.775 & 0.808 & 0.779 & 0.724  & \multicolumn{1}{l|}{0.932} \\
 & \textit{fea-3} & 0.778 & 0.808 & 0.780 & 0.726 &  \multicolumn{1}{l|}{0.932} \\ \hline
\multirow{3}{*}{$MLP_{deep}$} & \textit{fea-1} & 0.633  & 0.743  & 0.653  & 0.625  & \multicolumn{1}{l|}{0.945} \\
 & \textit{fea-2} & 0.801 & 0.830  & 0.805   & 0.769 & \multicolumn{1}{l|}{0.921} \\
 & \textit{fea-3} & 0.815 & 0.831 & 0.816 & 0.778 & \multicolumn{1}{l|}{0.923} \\ \hline
$\ourss_{shallow}$ & / & 0.905 & 0.929 & 0.908 & 0.905 & \multicolumn{1}{l|}{0.933} \\
$\ourss_{deep}$ & / & \textbf{0.929} & \textbf{0.961} & \textbf{0.935} & \textbf{0.932} & \multicolumn{1}{l|}{0.936} \\ \hline
\end{tabular}
\caption{Comparison on normal program reidentification.}
\label{tab:overall}
\vspace{-12pt}
\end{table*}
In this experiment, we evaluate the performance on normal program reidentification by comparing \ourss\ with other four baselines including LR, SVM, XGB, and MLP. We use the constructed features as described in Section~\ref{subsec:data}. %Based on these features, we consider four typical classification models, i.e., Logistical Regression, SVM, XGB and MLP. Among them XGB is the most powerful and scalable classification model. MLP is a typical neural network based model. 
To show the effectiveness of the deep learning model, we consider both shallow and deep versions of the neural network based model, which is denoted as $"XXX_{shallow}"$ and $"XXX_{deep}"$, respectively. We consider the one-layer configuration as the shallow model and the three-layer configuration as the deep model. For the baseline methods, we use the constructed raw features: \textit{fea-1}, \textit{fea-2}, or \textit{fea-3} as the input, respectively.
Table \ref{tab:overall} shows that overall, \ourss\ consistently and significantly outperforms all baselines in terms of all metrics. More specifically, (1) \textit{fea-3} (concatenate the connectivity feature and graph statistic feature) helps to improve the performance for all baseline classification models, but since they are all raw features without considering graph structure, their performance can not catch up with the proposed method; (2) \textit{fea-2} (graph statistic feature) is more useful than \textit{fea-1} (graph connectivity feature), since \textit{fea-1} is very sparse, which is difficult to use for some state-of-art methods, such as SVM; (3) The deep model outperforms the shallow model for our approach, since the deep model can capture the hierarchical dependency representation. 
\begin{figure}[htb]
\vspace{-8pt}
\centering
\includegraphics[width = 0.8\linewidth]{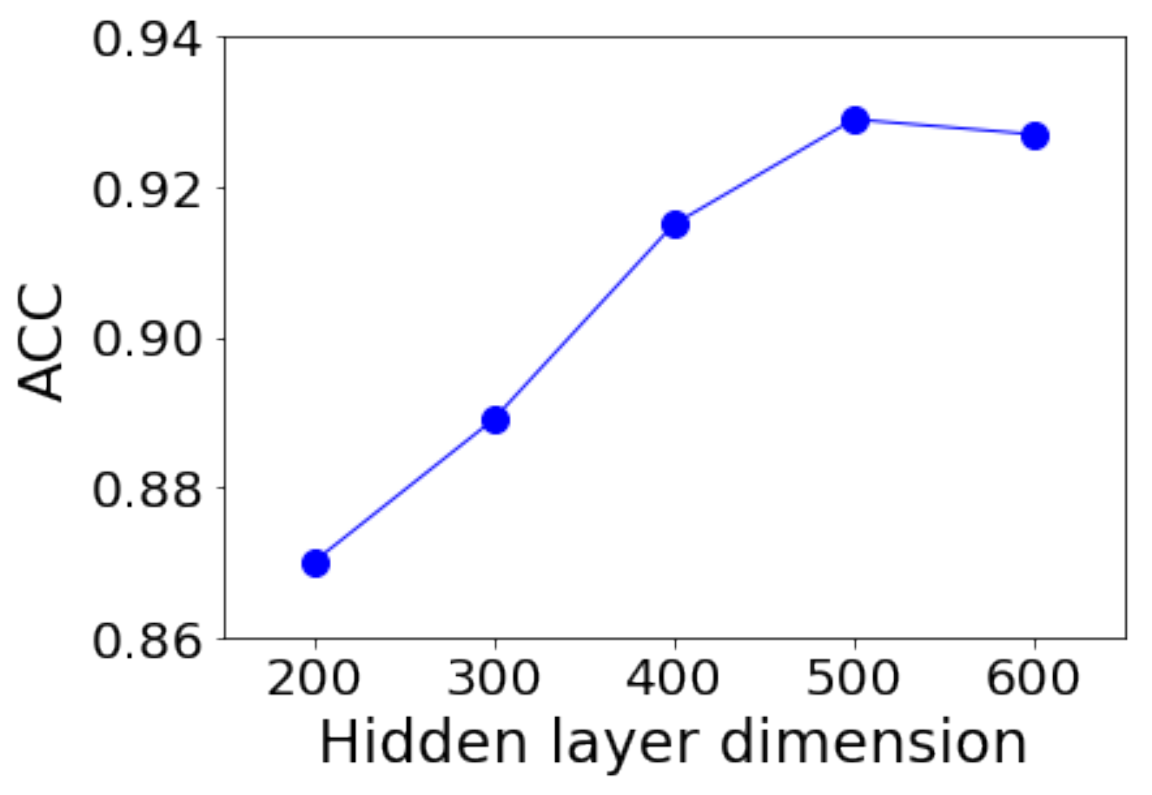}
 \vspace{-12pt}
\caption{Parameter sensitivity analysis results.}
\label{fig:p_dim}
\vspace{-8pt}
\end{figure}
% \vspace{-5pt}
% \subsubsection{Parameter Sensitivity}
% \begin{figure}[htb]
% \vspace{-5pt}
% \centering
%  \subfigure[\# contexts vs. ACC]{\label{fig:p_cont}
%     \begin{minipage}[l]{0.45\linewidth}
%      \centering
%      \includegraphics[width=\linewidth]{fig/P_CON.pdf}
%     \end{minipage}
%  }~
%  \subfigure[\# dimensions vs. ACC]{\label{fig:p_dim}
%     \begin{minipage}[l]{0.45\linewidth}
%      \centering
%      \includegraphics[width=\linewidth]{fig/P_DIM.pdf}
%     \end{minipage}
%  }
% \vspace{-8pt}
% \caption{Parameter sensitivity evaluation.}\label{fig:exp_comp1}
% \vspace{-10pt}
% \end{figure}

In this experiment, we also conduct the sensitivity analysis of how different choices of parameters will affect the performance of our proposed approach. Specifically, \ourss\ has an important parameter: the number of dimensions in each perceptron layer. Figure \ref{fig:p_dim} shows that the performance of \ourss\ is not significantly affected by the number of hidden layer dimensions.\nop{The results are %shown in Figure \ref{fig:p_cont} and \ref{fig:p_dim}.
shown in Figure \ref{fig:p_dim}.}

\vspace{-5pt}
\subsection{Real-world Experiments}
In this section, we evaluate the performance of the \ourss\ in two real-world tasks: disguised program detection and normal program upgrade detection.
\vspace{-5pt}
\subsubsection{Disguised Program Detection}
\begin{figure}[htb]
\vspace{-10pt}
\centering
 \subfigure[TPR Results.]{\label{fig:real1_TPR}
    \begin{minipage}[l]{0.45\linewidth}
     \centering
     \includegraphics[width=\linewidth]{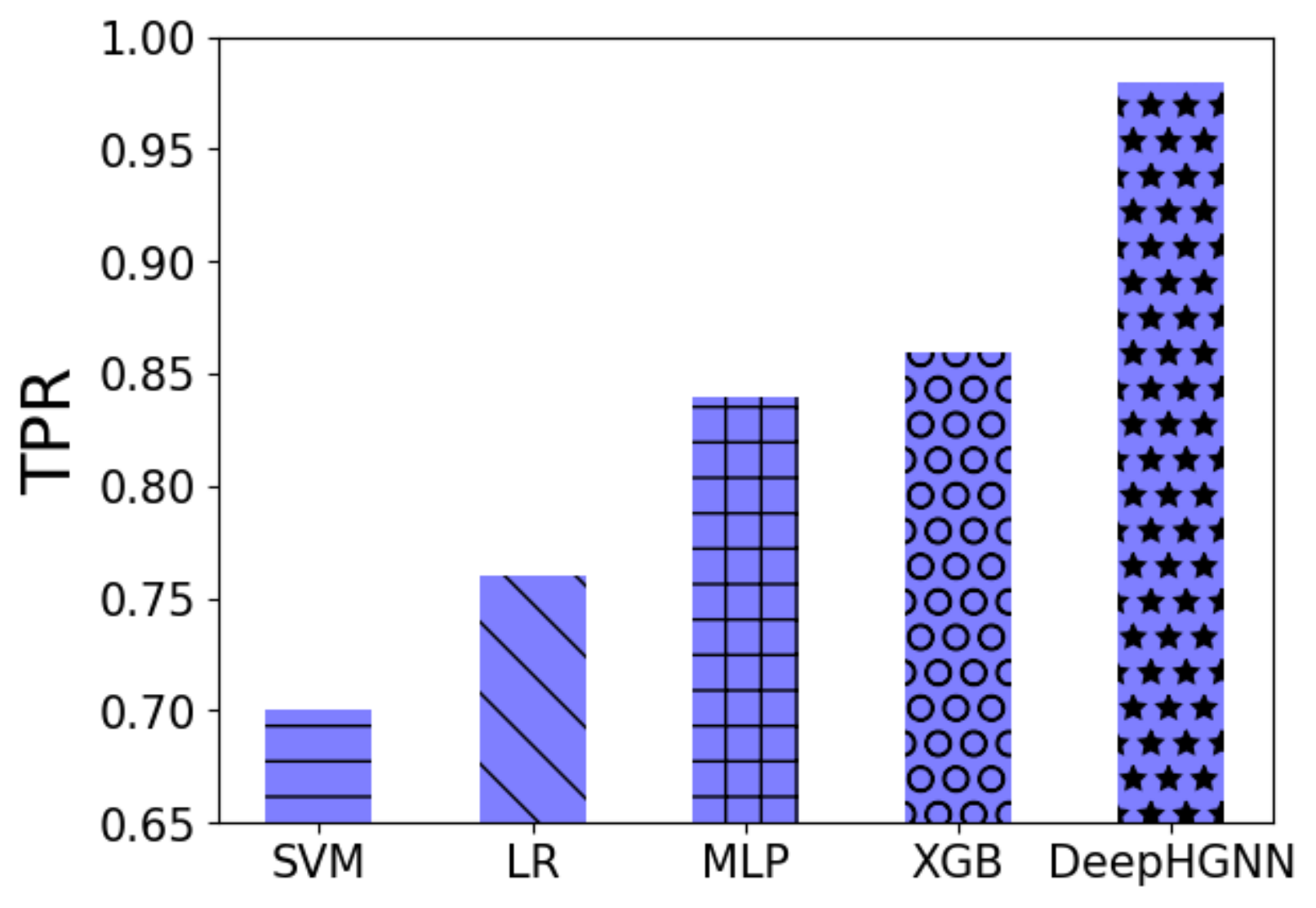}
    \end{minipage}
 }~
 \subfigure[FPR Results.]{\label{fig:real2_FPR}
    \begin{minipage}[l]{0.45\linewidth}
     \centering
     \includegraphics[width=\linewidth]{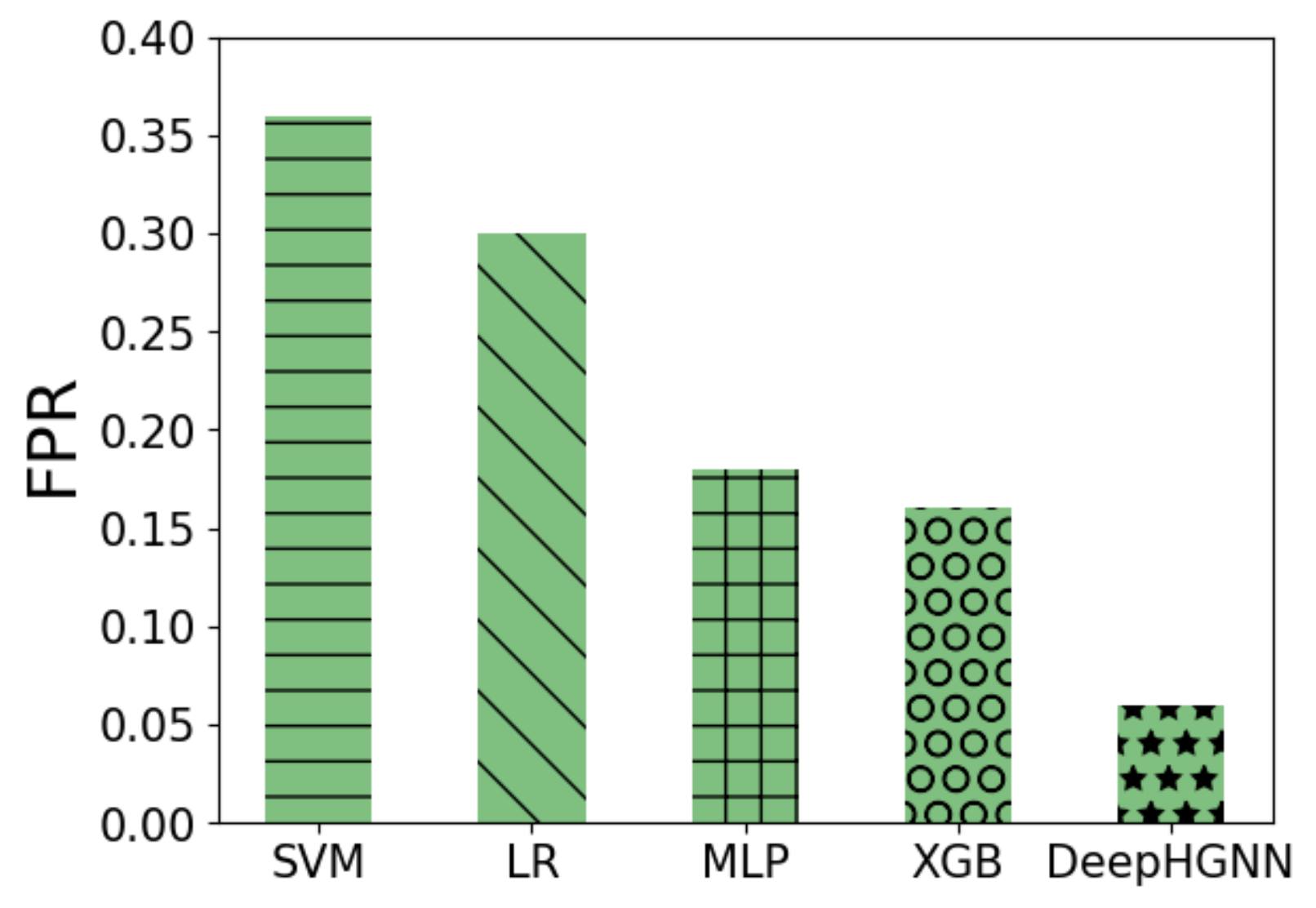}
    \end{minipage}
 }
 \vspace{-10pt}
\caption{Disguised program detection results.}\label{fig:real1}
\vspace{-5pt}
\end{figure}

In order to avoid detection, hackers usually disguise their malicious program as a legitimate program. To test the performance of \ourss\ in disguised program detection, %in this experiment, %we use \ourss\ in a real-world disguised program detection task.
we simulate five different types of real advanced attacks including: 
(1) \textit{WannaCry}: A crypto worm which disguised itself as 7z.exe;
(2) \textit{Phishing Email}: A malicious Trojan is downloaded as an Outlook attachment, and the enclosed macro is triggered by Excel to create a fake java.exe, and the disguised malicious java.exe further exploits a vulnerable server to start cmd.exe in order to create an info-stealer;
(3) \textit{Emulating Enterprise Environment}: The hackers generate telnet process to create a Trojan malware binary with a disguised normal program name. Then, %Trojan is created to connect back to hackers, and 
DLL is injected through the running process notepad.exe. The hackers use mimikaz and kiwi for memory operation inside the meterpreter context. Finally, malware PwDump7.exe and wce.exe are copied and run on target hosts;
(4) \textit{Diversifying Attack Vectors}: The hacker first writes malicious PHP file by HTTP connection, then downloads the malware process (Trojan.exe), and connects back to the hacker host. The process notepad.exe is run to perform DLL injection. The hacker further uses mimikaza and kiwi to perform memory operation inside meterpreter context. Finally, it copies and runs Pwdup7.exe and wce.exe on the target host;
(5) \textit{Domain Controller Penetration}: The hackers first send an email attaching a document that includes the malware python32.exe. This malware opens a connection back to the hackers so that they can run notepad.exe and perform reflective DLL injection to obtain the needed privileges. Finally, they transfer password enumerator and run the process gsecdumpv2b5.exe to get all user credentials. %Finally, SQL server address is probed to connect and dump the database into the attacker host.
%they probe the SQL server address and dump the database into their own bases. 
We try each type of the attacks ten times during different time windows to generate different testing samples.

%In our experiment, it detects the disguised program by identifying if its behavior matches its claimed name. 
The detection performance is evaluated using the true positive rate (TPR) and false positive rate (FPR). 
The TPR defines the fraction of intrusion attacks that are detected during the evaluation. FPR, on the other hand, describes the fraction of normal event sequences that trigger an alert during the evaluation. We compare our method with the four baselines and use the deep model for all neural network based techniques. Figure \ref{fig:real1} shows that \ourss\ outperforms all the other baselines by at least $12\%$ in TPR and $10\%$ in FPR.% in recall
\subsubsection{Robustness Study on Program Upgrade}
\begin{figure}[htb]
\vspace{-12pt}
\centering
\includegraphics[width = 0.9\linewidth]{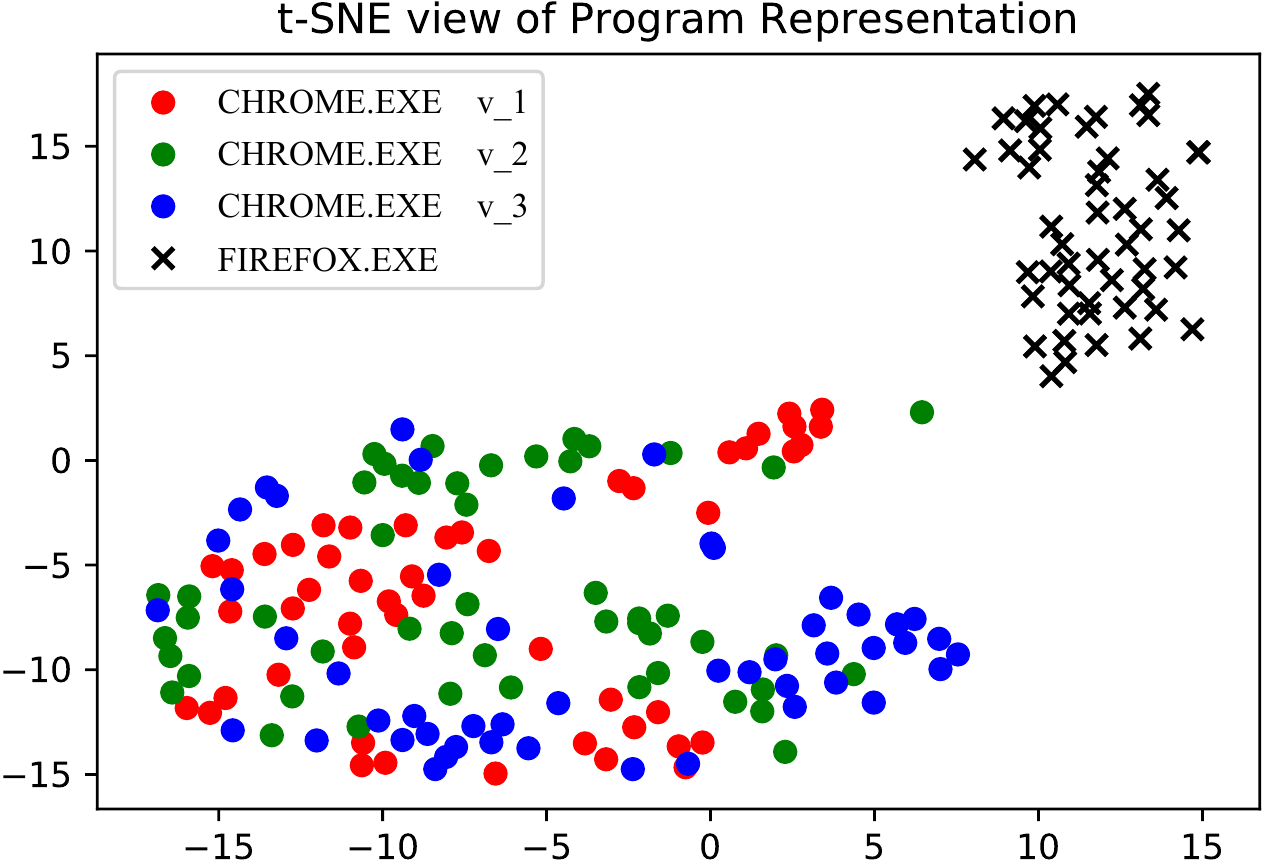}
 \vspace{-3pt}
\caption{Scatter plot embedding of different versions of CHROME.EXE vs FIREFOX.EXE.}
\label{fig:t-sne}
\vspace{-3pt}
\end{figure}
The normal program usually gets the upgrade with multiple versions. However, different versions have different signatures, which may cause false alarms in intrusion detection systems.
In this experiment, we evaluate the robustness of \ourss\ on normal program upgrades. In particular, we select the web browser software CHROME.EXE for a case study. We collect the data of CHROME.EXE with three different versions and one version of another web browser software FIREFOX.EXE. Then, the graph embedding before the classifier for these CHROME.EXE data and FIREFOX.EXE data are extracted and visualized using t-SNE. From Figure \ref{fig:t-sne}, we can see that in the embedding space, the data points belong to the different versions of CHROME.EXE are close to each other, even though they have different signatures, while the data points of FIREFOX.EXE are far away from all the CHROME.EXE data points, which demonstrates our method is robust to the normal program upgrades.

\vspace{-6pt}
\section{Related Work}
%Our work is related to Intrusion Detection and Graph Representation Learning. We briefly discuss them in the followings.

\subsection{Intrusion Detection}
\nop{
The malicious program detection powered by machine Learning and data mining is an emerging research topic due to the increasing number of malicious programs and the developing the of the AI algorithm. There are a number work are proposed based on the machine learning and data mining \cite{arp2014drebin,dimjavsevic2016evaluation,wu2014droiddolphin,saleh2017multi,hou2017hindroid}. These work follow a two-stage pipeline and formulate the malicious program detection as a binary classification problem. They first extract the feature from the malware samples and then perform binary classification using different classifier, such as K-NN, Naive Bayesian, Decision Tree and SVM. More recently, a number of work adopt the deep learning to perform this pipeline end-to-end. \cite{saxe2015deep} proposes to use Auto-Encoder and Deep Belief Network to extract hidden representation as the signature of the malware. 
\cite{david2015deepsign} combines the deep neural network and Bayesian Calibrate Model.
\cite{hardy2016dl4md} proposes SAEs based deep learning architecture to learn from Windows API calls for malware detection.
Different from previous work, our work is in a general view, which has broader applications, including Malicious Program Detection. 
}

In general, intrusion detection refers to the process of monitoring the events occurring in a computer system or network and analyzing them for signs of intrusions. Currently, there are two main types of approaches, namely anomaly detection and misuse detection \cite{jones2000computer}. Anomaly detection approaches define and characterize correct/wrong behaviors of the system, while the misuse detection approaches monitor for explicit patterns, with the intrusion patterns known in advance.
%\noindent \textbf{Anomaly Detection.} 
Existing anomaly detection approaches in large-scale enterprise network systems have been separately considering different data representations. In particular, host-based anomaly detection methods \cite{DongCWT0LLC17,CaoCCTLL18,LinCCT0CL18} locally extract patterns from process-level events as the discriminators of abnormal intrusion. In contrast, network-based anomaly detection methods \cite{noble2003graph} focus on disclosing abnormal subgraph structures from network-level events, most of which are inspired by graph properties.%\cite{ide2004eigenspace,noble2003graph}. %All these related work, however, did not take into account both types of events together. Different from these work, we consider the heterogeneity among data.
\vspace{-5pt}
\subsection{Graph Representation Learning}
Representation learning \cite{bengio2013representation} has become a very promising topic in machine learning with wide applicability. %Recently representation learning \cite{bengio2013representation} has long been an important problem of machine learning, and 
Many representation learning work aim at learning representations with deep learning due to its powerful feature learning ability. In recent years, several deep learning approaches have been proposed on graph-structured data, including \cite{kipf2016semi,defferrard2016convolutional,hamilton2017inductive,velickovic2017graph}. They leverage convolution operation in the spatial domain or spectral domain. \cite{defferrard2016convolutional} proposes the fast localized convolutions. \cite{kipf2016semi} proposes a first-order approximation scheme to reduce the computation cost of graph filter spectrum. \cite{hamilton2017inductive} extends the graph convolution with a more general form of sample and aggregation function for node context. More recent work \cite{velickovic2017graph} learns the context aggregation with considering the weights between the neighborhood and current nodes. 
However, these work mainly focus on the node classification for heterogeneous network. Different from theirs, our work focuses on learning the embedding from the heterogeneous graph. 
\vspace{-6pt}
\section{Conclusion}
\nop{
In this paper, we propose the problem of program verification via deep graph neural network in a data driven manner.
\myred{Rewrite the conclusion. Conclusion should be totally different to the abstract.}
In this paper, we propose a graph neural network based method (\ourss) to verify the program based on its monitoring and surveillance data. The key idea is to leverage the representation learning of the program behavior graph to guide the verification process. We formulate this problem from a graph classification perspective, and develop an effective attentional heterogeneous graph embedding algorithm to solve it. Extensive experiments --- using real-world enterprise monitoring data and real attacks --- show that \ourss\ is effective in identifying the disguised signed programs and robust to the normal dynamic changes like program version upgrade.
}

In this paper, we investigate the problem of program reidentification in IT/OT systems, which is often overlooked by traditional intrusion detection techniques. We propose \ourss, an attentional heterogeneous graph neural network method to verify the program's identity based on its heterogeneous behavior graph. %\ourss\ is the first deep learning framework for graph classification task on heterogeneous directed graphs. 
%We address the challenge to leverage the system monitoring data. 
Different from the existing homogeneous graph neural network methods, \ourss\ is able to capture and encode the heterogeneous complex dependency among different entities in a hierarchical way. The experimental results show that \ourss\ outperforms all the baseline methods by at least $10\%$ in terms of all the metrics. 
We also apply \ourss\ to a real enterprise system for disguised program detection. Our method can achieve superior performance and demonstrate robustness across the normal dynamic changes. 

%The real-world experiments show that \ourss\ the effectiveness in identifying the disguised signed programs and robust to the normal dynamic changes like program version upgrade.    

% to . A multi-channel transformation module is developed to translate the heterogeneous graph to multiple channels of homogeneous graph. A contextual graph encoder is then proposed to extract the graph embedding preserving the directed dependency. A channel-aware attention module is designed to align the graph embedding from different channels. 
% We implement and deploy our approach to a real enterprise security system, and evaluate the proposed algorithm
% in extensive experiments. The experiment results convince us of the effectiveness and efficiency of our approach.

\vspace{-6pt}
\section{Acknowledgements}
The first author is supported in part by NSF through grants IIS-1526499, IIS-1763325, and NS-1626432.
%, and NSFC 61672313.
%\bibliographystyle{plain}
%\myred{Check the format and consistence of the references}


\vspace{-6pt}
\begin{thebibliography}{10}
\providecommand{\url}[1]{\texttt{#1}}
\providecommand{\urlprefix}{URL }
\vspace{-6pt}
\bibitem{kernelaudit}
\url{https://wiki.archlinux.org/index.php/Audit_framework} (2016)

\bibitem{etw}
\url{https://msdn.microsoft.com/en-us/library/ff357719} (2017)

\bibitem{arefi2018faros}
Arefi, M.N., Alexander, G., \textit{et al.}: Faros: illuminating in-memory
  injection attacks via provenance-based whole-system dynamic information flow
  tracking. In: 48th Annual IEEE/IFIP International Conference on Dependable
  Systems and Networks. pp. 231--242 (2018)

\bibitem{atwood2016diffusion}
Atwood, J., Towsley, D.: Diffusion-convolutional neural networks. In: Advances
  in Neural Information Processing Systems. pp. 1993--2001 (2016)

\bibitem{bengio2013representation}
Bengio, Y., Courville, A., Vincent, P.: Representation learning: A review and
  new perspectives. IEEE Transactions on Pattern Analysis and Machine
  Intelligence  35(8),  1798--1828 (2013)

\bibitem{CaoCCTLL18}
Cao, C., Chen, Z., \textit{et al.}: Behavior-based community detection:
  Application to host assessment in enterprise information networks. In: 27th
  {ACM} International Conference on Information and Knowledge Management. pp.
  1977--1985 (2018)

\bibitem{chen2016xgboost}
Chen, T., Guestrin, C.: Xgboost: A scalable tree boosting system. In: 22nd ACM
  SIGKDD International Conference on Knowledge Discovery and Data Mining. pp.
  785--794. ACM (2016)

\bibitem{chen2017task}
Chen, T., Sun, Y.: Task-guided and path-augmented heterogeneous network
  embedding for author identification. In: Tenth ACM International Conference
  on Web Search and Data Mining. pp. 295--304 (2017)

\bibitem{defferrard2016convolutional}
Defferrard, M., Bresson, X., \textit{et al.}: Convolutional neural networks on
  graphs with fast localized spectral filtering. In: Advances in Neural
  Information Processing Systems. pp. 3844--3852 (2016)

\bibitem{DongCWT0LLC17}
Dong, B., Chen, Z., \textit{et al.}: Efficient discovery of abnormal event
  sequences in enterprise security systems. In: 26th {ACM} International
  Conference on Information and Knowledge Management. pp. 707--715 (2017)

\bibitem{gao2018saql}
Gao, P., Xiao, X., \textit{et al.}: Saql: a stream-based query system for
  real-time abnormal system behavior detection. arXiv preprint arXiv:1806.09339
   (2018)

\bibitem{hamilton2017inductive}
Hamilton, W., Ying, Z., \textit{et al.}: Inductive representation learning on
  large graphs. In: Advances in Neural Information Processing Systems. pp.
  1024--1034 (2017)

\bibitem{pki}
Hendric, W.: A complete overview of trusted certificates. In: CA/Browser Forum
  (2015)

\bibitem{jones2000computer}
Jones, A.K., Sielken, R.S.: Computer system intrusion detection: a survey.
  Computer Science Technical Report pp. 1--25 (2000)

\bibitem{kipf2016semi}
Kipf, T.N., Welling, M.: Semi-supervised classification with graph
  convolutional networks. arXiv preprint arXiv:1609.02907  (2016)

\bibitem{LinCCT0CL18}
Lin, Y., Chen, Z., \textit{et al.}: Collaborative alert ranking for anomaly
  detection. In: 27th {ACM} International Conference on Information and
  Knowledge Management. pp. 1987--1995 (2018)

\bibitem{liu2018towards}
Liu, Y., Zhang, M., \textit{et al.}: Towards a timely causality analysis for
  enterprise security. In: Network and Distributed Systems Security Symposium
  (2018)

\bibitem{noble2003graph}
Noble, C.C., Cook, D.J.: Graph-based anomaly detection. In: ACM International
  Conference on Knowledge Discovery and Data Mining. pp. 631--636 (2003)

\bibitem{ren2014droidmarking}
Ren, C., Chen, K., Liu, P.: Droidmarking: resilient software watermarking for
  impeding android application repackaging. In: 29th ACM/IEEE International
  Conference on Automated Software Engineering. pp. 635--646 (2014)

\bibitem{sun2011pathsim}
Sun, Y., Han, J., \textit{et al.}: Pathsim: Meta path-based top-k similarity
  search in heterogeneous information networks. VLDB Endowment  (2011)

\bibitem{tang2018node}
Tang, Y., Li, D., \textit{et al.}: Nodemerge: template based efficient data
  reduction for big-data causality analysis. In: 25th ACM Conference on
  Computer and Communications Security (2018)

\bibitem{velickovic2017graph}
Velickovic, P., Cucurull, G., \textit{et al.}: Graph attention networks. arXiv
  preprint arXiv:1710.10903  (2017)

\bibitem{verleysen2005curse}
Verleysen, M., Fran{\c{c}}ois, D.: The curse of dimensionality in data mining
  and time series prediction. In: International Work-Conference on Artificial
  Neural Networks. pp. 758--770. Springer (2005)

\bibitem{zhang2015categorical}
Zhang, K., Wang, Q., \textit{et al.}: From categorical to numerical: Multiple
  transitive distance learning and embedding. In: 2015 SIAM International
  Conference on Data Mining. pp. 46--54 (2015)

\bibitem{zimba2017malware}
Zimba, A.: Malware-free intrusion: a novel approach to ransomware infection
  vectors. International Journal of Computer Science and Information Security
  15(2),  317 (2017)

\end{thebibliography}
\end{document}